\documentclass[a4paper,fleqn]{cas-dc}
\graphicspath{ {./figures/} }

\usepackage[authoryear,longnamesfirst]{natbib}
\usepackage{amsmath,amsbsy,amsmath,amsthm,amssymb,amsfonts}
\usepackage{graphicx,xcolor}
\usepackage{booktabs}
\usepackage{bm}
\usepackage{graphicx}
\usepackage{algorithmic,float}
\usepackage{array}
\usepackage[caption=false,font=normalsize,labelfont=sf,textfont=sf]{subfig}
\usepackage{textcomp}
\usepackage{stfloats}
\usepackage{url}
\usepackage{verbatim}
\usepackage{graphicx}
\usepackage[ruled,vlined, linesnumbered]{algorithm2e}
\usepackage{booktabs}
\usepackage{orcidlink}
\usepackage{enumerate}
\usepackage{epsfig}
\usepackage{rotating}
\usepackage{csvsimple}
\usepackage{lscape}
\usepackage{adjustbox}
\usepackage{titletoc}
\usepackage{longtable}
\usepackage{tabularx}
\usepackage{xtab}
\usepackage{afterpage}
\usepackage[T1]{fontenc}
\usepackage{mwe}  
\hypersetup{
	pdfsubject={Online learning with radial basis function networks},
	pdfkeywords={online learning, feature representation transfer, financial time series, covariate shift, concept drift},
	pdfauthor={Gabriel Borrageiro},
	pdftitle={Online learning with radial basis function networks},
}

\def\tsc#1{\csdef{#1}{\textsc{\lowercase{#1}}\xspace}}
\tsc{WGM}
\tsc{QE}

\begin{document}
	\let\WriteBookmarks\relax
	\def\floatpagepagefraction{1}
	\def\textpagefraction{.001}
	
	\shorttitle{Online learning with radial basis function networks}    
	\shortauthors{Borrageiro et al.}  
	
	\title{Online learning with radial basis function networks}  
	
	\author[1]{Gabriel Borrageiro}[orcid=0000-0002-0063-7103]
	\cormark[1]
	\ead{gabriel.borrageiro.20@ucl.ac.uk}
	
	\affiliation[1]{organization={University College London},
		addressline={Gower Street}, 
		city={London},
		postcode={WC1E 6BT}, 
		country={United Kingdom}}
	
	\author[1]{Nick Firoozye}[orcid=0000-0002-6460-0406]
	\author[1]{Paolo Barucca}[orcid=0000-0003-4588-667X]
	
	\cortext[1]{Corresponding author}
	
	\begin{abstract}
		Financial time series are characterised by their nonstationarity and autocorrelation. Even if these time series are differenced, technically ensuring their stationarity, they experience regular covariate shifts and concept drifts. Against this backdrop, we combine feature representation transfer with sequential optimisation to provide multi-horizon returns forecasts. Our online learning rbfnet outperforms a random-walk baseline and several powerful batch learners. The rbfnets we formulate are naturally designed to measure the similarity between test samples and continuously updated prototypes that capture the characteristics of the feature space.
	\end{abstract}
	
	
	\maketitle
	
	\section{Introduction} \label{sec:intro}
	Financial time series provide several modelling challenges and are typically both serially correlated and nonstationary. The dynamics of financial assets have been modelled as a jump-diffusion process \citep{Merton76}, which is now commonly used in econometrics. The jump-diffusion process implies that financial time series should observe small, continuous changes and occasional jumps over time. \cite{Bachelier1900} considers price series as Gaussian random walks, whose increments are iid Gaussian random variables. Bachelier's first law states that the variation of returns grows with the square root of time. \cite{Bouchaud2018} find that Bachelier's first law holds well for actual financial returns; however, they also find that standard Gaussian random-walk models for financial returns modelling underestimate the extreme fluctuations that are empirically observed. Furthermore, they find that price changes follow fat-tailed, power-law distributions, with extreme events not as rare as Gaussian models might predict. Finally, they find that market activity and volatility are highly intermittent in time, with intense activity intertwined with periods of calm; these are the behavioural characteristics modelled by the jump-diffusion hypothesis. 
	
	One approach for coping with nonstationarity is to learn online continuously. Sequential model fitting may be combined with states-of-nature/transitional learning approaches such as reinforcement learning or continual learning approaches such as transfer learning \citep{yang_zhang_dai_pan_2020}. We combine feature representation transfer with sequential optimisation, and the result is an online learning radial basis function network (rbfnet). The rbfnet benefits from feature representation transfer from clustering algorithms that determine the network's structure. When providing multi-horizon returns forecasts for the major cross-asset class time series, our rbfnet obtains the best test-set results with minimum mean squared error. Our rbfnet outperforms a random-walk baseline and several powerful batch learners.
	
	\section{The radial basis function network} \label{sec:rbfnet}
	The rbfnet is a single-layer network whose hidden units are radial basis functions (rbf) of the form
	\begin{equation} \label{eq:radial_basis_function}
		\phi_j(\mathbf{x}) = \exp{\left( -\frac{1}{2}[\mathbf{x} - \bm{\mu}_j]^T \bm{\Sigma}_j^{-1} [\mathbf{x} - \bm{\mu}_j] \right)}.
	\end{equation}
	The hidden unit means and covariances are typically learnt through clustering algorithms such as k-means \citep{LloydS1982Lsqi}. The hidden unit outputs are aggregated into a feature vector 
	\begin{equation}
		\bm{\phi}_t = [1, \phi_1(\mathbf{x}), ..., \phi_k(\mathbf{x})],
	\end{equation}
	and mapped to the response via regression
	\begin{equation}
		y_t = \bm{\theta}^T \bm{\phi}_t + \epsilon.
	\end{equation}
	The rbfnet is amenable to sequential optimisation via exponentially weighted recursive least squares (ewrls) \citep{KernelAdaptiveFiltering2010}.
	
	A multilayer perceptron separates classes using hidden units that form hyperplanes in the input space. The separation of class distributions modelled by local radial basis functions is probabilistic. The predictive uncertainty increases where there is class-conditional distribution overlap. The radial basis function (rbf) activations can be thought of as the posterior feature probabilities, and the weights can be interpreted as the posterior probabilities of class membership, given the presence of the features \citep{BishopChristopherM1995Nnfp}.
	
	\begin{figure}[htpb]
		\centering
		\caption{The radial basis function network}
		\label{fig:rbfnet}
		\includegraphics[width=.9\columnwidth]{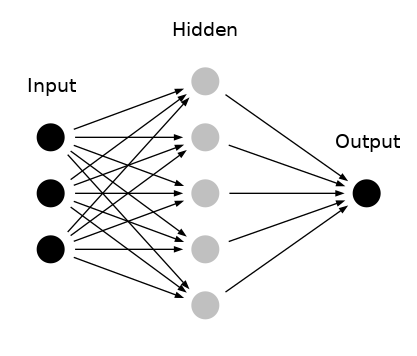}
	\end{figure}

	\section{The research experiment}
	We consider the goal of multi-step forecasting with financial time series returns. Define the prediction mean squared error (mse) for the $j'th$ model and $h'th$ forecast horizon as
	\begin{equation}
		mse_{h, j} = \frac{1}{t-h} \sum_{i=1}^{t-h} (y_{i+h, j} - \hat{y}_{i, j})^2.
	\end{equation}
	The performance criteria that we consider is normalised mse for forecast horizons in days $h=1, ..., 30$
	\begin{equation} \label{eq:nmse}
		nmse_{h, j} = mse_{h, j} / mse_{rw}.
	\end{equation}
	In equation \ref{eq:nmse}, the normalisation of mse occurs relative to the random-walk baseline. The random-walk \citep{Harvey1993}
	\begin{equation}
		y_t = y_{t-1} +\epsilon_t,
	\end{equation}
	has stationary expectation
	\begin{equation} \label{eq:rw_expectation}
		\mathbb{E}[y_t] = y_0,
	\end{equation}
	and nonstationary variance/covariance
	\begin{align}
		Var[y_t] &= t\sigma^2 \\
		Cov[y_t, y_{t-\tau}] &= |t-\tau]\sigma^2.
	\end{align}
	A large body of academic literature shows that it is difficult to beat the random-walk model when forecasting returns of financial time series \citep{MeeseRichardA1983Eerm, EngelCharles1994CtMs}. If $nmse_{h, j} < 1$, then we conclude that model $j$ has outperformed the random-walk for forecast horizon $h$.
	
	\subsection{The Refinitiv dataset} \label{sec:dataset_rbfnet}
	Refinitiv is a global financial market data and infrastructure provider. We use their Data Platform Python package to extract daily-sampled data, including currency pairs, equities, rates, credit, metals, commodities, energy and crypto. Refinitiv extracts the datasets from multiple exchanges, electronic communication networks and over-the-counter market makers. The sampled prices are the last daily traded or quoted limit order book price, with a snapshot taken at 5 pm EST. The dataset begins on 2018-11-01 and ends on 2022-05-20. We reserve half the data for training (649 observations) and the remaining half for testing (648 observations). 
	
	\subsection{Feature selection} \label{sec:feat_select}
	There are various feature selection algorithms, which \cite{ESL} discuss in detail. Forward stepwise selection scales well as the dimensionality $d$ of the feature space $\mathbf{X} \in \mathbb{R}^{n \times d}$ increases. The goal of forward stepwise selection is to choose features that maximise $R^2$, although the features might be positively correlated. On the other hand, variance inflation factor (vif) minimisation \citep{JamesGarethGarethMichael2013Aits} performs feature selection by minimising the correlation between features. We combine forward stepwise selection and vif minimisation to the training set returns. The feature selection is applied to each target, and the target-conditional subset of external inputs is held fixed during the test set evaluation.

	\subsection{Competitor models} \label{sec:competitor_models}
	We consider several competitor models in our experiment. We use scikit-learn \citep{scikit-learn} implementations for Gaussian process regression (gpr), gradient tree boosting (gtb), k-nearest neighbours regression (knn), the multilayer perceptron (mlp), the random forest (rf) and support vector regression (svm). We use our implementations of ridge regression (ridge), kernel ridge regression (k-ridge), ewrls and the rbfnet. The online rbfnet faces a robust assortment of competitor models, which in most cases are suited to or can only be fitted by batch learning. 
	
	\subsection{Experiment design} \label{sec:experiment_design}
	We construct daily returns and set half the data aside for training and the other half for testing. We perform external input feature selection as per section \ref{sec:feat_select} on the training set returns (per target), which are held fixed in the test set. Hyperparameters are set ex-ante. For example, we use the scikit-learn defaults and our rbfnet, and ewrls models use an exponential decay factor of $\tau=0.99$. During test time, these two models continue to be fitted online. We consider a second performance measure, forecast accuracy
	\begin{equation}
		acc_{h, j} = \frac{1}{t-h} \sum_{i=1}^{t-h} I[sign(y_{i+h, j}) = sign(\hat{y}_{i, j})],
	\end{equation}
	where $I[.]$ is an indicator function that returns 1 for a true condition, or else 0, and
	\begin{equation}
		sign(x) =
		\begin{cases}
			1 & \text{if $x > 0$} \\
			0 & \text{if $x = 0$} \\
			-1 & \text{if $x < 0$} \\
		\end{cases}.
	\end{equation}
	
	\section{Results} \label{sec:rbfnet_results}
	Table \ref{tab:rbfnet_test_set_nmse_by_model} shows that several models have average test set nmse that is better than the random-walk baseline. These include ewrls, gpr, gtb, k-ridge, rbfnet and rf. The models that perform worse than the random-walk baseline include knn, mlp, ridge and svm, with the mlp being the worst-performing model. The rbfnet has the lowest average nmse of 0.636. Comparing this to the second-best result, a nmse of 0.673 for gpr, we perform a two-sample t-test for equal means \citep{snedecor1989statistical} and find that the means are considered statistically different, drawn from differently parameterised distributions. For all models, we also perform a Wald test \citep{Wasserman2004} for the null hypothesis that the nmspe is no different from 1, tested at the 5\% critical value. We find that in all cases, the model-averaged nmse is statistically different from 1. 
	
\begin{figure}[htpb]
		\centering
		\caption{Test set nmse by model and forecast horizon in days.}
		\label{fig:rbfnet_nmse_markouts}
		\includegraphics[width=1.\columnwidth]{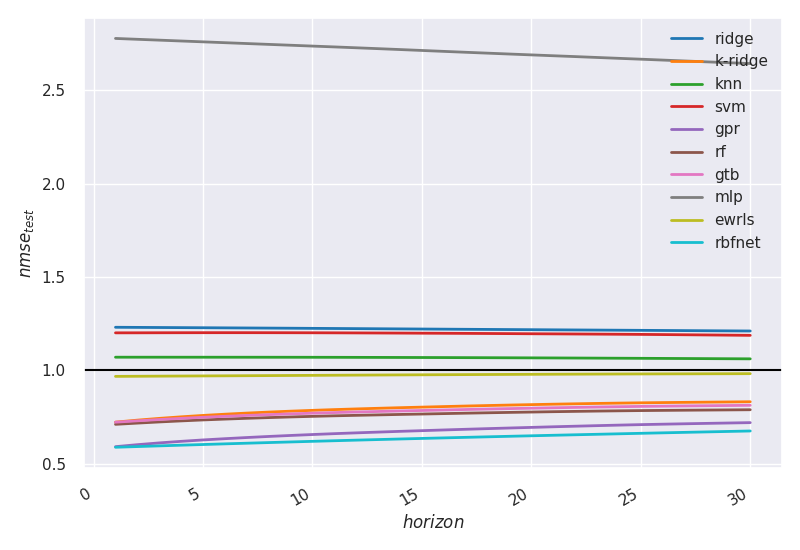}
	\end{figure}

	Figure \ref{fig:rbfnet_nmse_markouts} shows the nmse by model and forecast horizon in days. There is a similar performance between the rbfnet and gpr for $h=1$. For $h=2,...,30$, the rbfnet outperforms gpr, the random-walk baseline and the remaining competitor models. We cannot put the rbfnet outperformance down to sequential updating alone in the test set; if this were the case, the ewrls model would outperform the remaining offline learning models. Instead, ewrls performs worse than gpr, rf, gtb and k-ridge offline learning models.
	
	Table \ref{tab:rbfnet_test_set_accuracy_by_model} shows forecast accuracy by model for the horizons $h=1, ..., 30$. The gtb has the highest accuracy at 83.1\%, followed by gpr at 82.7\%. Rf, k-ridge and the rbfnet follow closely behind. The worst-performing model is ridge regression, with 54.7\% accuracy. Nevertheless, a two-sample t-test for equal means, where the comparison is made between the ridge model accuracy and a Bernoulli distributed mean of 50\% with a variance of 0.25\%, concludes that the ridge model has statistically better accuracy than pure chance. 
	
	\begin{table}[htpb]
		\caption{Test set nmse by model.}
		\label{tab:rbfnet_test_set_nmse_by_model}
		\vspace{0.1cm}
		\smaller
		\begin{tabularx}{1.\columnwidth}{lXXXXXXXXXX}
			\toprule
			model &    ewrls &      gpr &      gtb &  k-ridge &      knn &      mlp &   rbfnet &       rf &    ridge &      svm \\
			\midrule
			targets    &  100 &  100 &  100 &  100 &  100 &  100 &  100 &  100 &  100 &  100 \\
			count      & 3000 & 3000 & 3000 & 3000 & 3000 & 3000 & 3000 & 3000 & 3000 & 3000 \\
			mean       &    0.98 &    0.67 &    0.78 &    0.8 &    1.07 &    2.71 &    \textbf{0.63} &    0.76 &    1.22 &    1.2 \\
			std        &    0.52 &    0.56 &    0.89 &    0.93 &    0.91 &    4.19 &    0.42 &    0.82 &    0.84 &    1.61 \\
			min        &    0.27 &    0.04 &    0.04 &    0.04 &    0.13 &    0.08 &    0.14 &    0.04 &    0.48 &    0.12 \\
			25\%        &    0.61 &    0.32 &    0.31 &    0.30 &    0.42 &    0.71 &    0.36 &    0.32 &    0.83 &    0.43 \\
			50\%        &    0.86 &    0.48 &    0.51 &    0.55 &    0.81 &    1.31 &    0.51 &    0.5 &    1.02 &    0.68 \\
			75\%        &    1.15 &    0.85 &    0.9 &    0.87 &    1.33 &    3.06 &    0.8 &    0.868 &    1.36 &    1.19 \\
			max        &    3.4 &    3.37 &    5.7 &    5.34 &    5.01 &   25.7 &    2.46 &    5.73 &    8.33 &   11.9 \\
			se         &    0.01 &    0.01 &    0.02 &    0.02 &    0.02 &    0.08 &    0.01 &    0.02 &    0.02 &    0.03 \\
			t-value    &   -2.49 &  -31.9 &  -13.4 &  -11.9 &    4.06 &   22.3 &  -48.1 &  -15.8 &   14.3 &    6.7 \\
			p-value    &    0.01 &    0.00 &    0.00 &    0.00 &    0.00 &    0.00 &    0.00 &    0.00 &    0.00 &    0.00 \\
			\bottomrule
		\end{tabularx}
	\end{table}
	
	\begin{table}[htpb]
		\caption{\smaller To two decimal places, test set accuracy by model averaged across the forecast horizons $h=1, ..., 30$ days.}
		\label{tab:rbfnet_test_set_accuracy_by_model}
		\vspace{0.1cm}
		\smaller
		\begin{tabularx}{1.\columnwidth}{lXXXXXXXXXX}
			\toprule
			model &    ewrls &      gpr &      gtb &  k-ridge &      knn &      mlp &   rbfnet &       rf &    ridge &      svm \\
			\midrule
			targets &  100 &  100 &  100 &  100 &  100 &  100 &  100 &  100 &  100 &  100 \\
			count     & 3000 & 3000 & 3000 & 3000 & 3000 & 3000 & 3000 & 3000 & 3000 & 3000 \\
			mean      &    0.63 &    0.83 &    \textbf{0.83} &    0.82 &    0.7 &    0.6 &    0.81 &    0.83 &    0.55 &    0.77 \\
			std       &    0.3 &    0.19 &    0.19 &    0.19 &    0.29 &    0.3 &    0.16 &    0.19 &    0.37 &    0.22 \\
			min       &    0.0 &    0.08 &    0.05 &    0.19 &    0.0 &    0.0 &    0.11 &    0.01 &    0.0 &    0.06 \\
			25\%       &    0.43 &    0.78 &    0.81 &    0.76 &    0.53 &    0.31 &    0.75 &    0.78 &    0.13 &    0.66 \\
			50\%       &    0.75 &    0.90 &    0.90 &    0.9 &    0.8 &    0.68 &    0.85 &    0.9 &    0.69 &    0.86 \\
			75\%       &    0.88 &    0.95 &    0.94 &    0.94 &    0.94 &    0.87 &    0.91 &    0.94 &    0.93 &    0.94 \\
			max       &    0.98 &    1.0 &    1.0 &    1.0 &    0.99 &    0.99 &    1.0 &    1.0 &    1.0 &    1.0 \\
			\bottomrule
		\end{tabularx}
	\end{table}
	
	\section{Discussion} \label{sec:online_rbfnet_discussion}
	Our rbfnets apply sequentially adapted feature representation transfer from clustering algorithms to supervised learners. Online transfer learning is a relevant area of research for nonstationary time series. Although transfer learning is primarily concerned with transferring knowledge from a source domain to a target domain and may be used offline or online, an increasing number of papers focus on online transfer learning \citep{Zhao_OTL_2014, Bruno_OTL_2019, Wang_OTL_2020}. Our paper contributes to continual learning in financial time series research by demonstrating that continual learning benefits multi-step forecasting, above and beyond sequential learning. If we compare the local learning of the rbfnet with the global learning technique of the feed-forward neural network, the latter suffers from catastrophic forgetting. \cite{Kirkpatrick_CF_2017} and \cite{Sukhov_CF_2020} look at ways of improving this issue, specifically at training networks that can maintain expertise on tasks that they have not experienced for a long time. The rbfnets we formulate are naturally designed to measure the similarity between test samples and continuously updated prototypes that capture the characteristics of the feature space. As such, the models are robust in mitigating catastrophic forgetting. 
	
	\section{Conclusion}
	Financial time series exhibit the attributes of autocorrelation, nonstationarity and nonlinearity. Our experiment demonstrates the added value of feature selection, nonlinear modelling, and online learning when providing multi-horizon forecasts. Technically, by constructing returns, the time series become stationary as measured by unit root tests; therefore, offline batch learning is possible. However, we find experimental evidence to support using sequentially optimised radial basis function networks (rbfnets) that utilise feature representation transfer. In addition, the rbfnets obtain the best experiment results, which can be attributed primarily to their clustering algorithms, which learn the intrinsic nature of the feature space. Finally, the resulting hidden units provide predictive prototypes for unseen test data, which retain high similarity across time.

	\bibliographystyle{apalike}
	\bibliography{main}
	
\end{document}